\pdfoutput=1 
\documentclass[preprint,12pt, number]{elsarticle}

\usepackage{amssymb}
\usepackage{siunitx}
\usepackage{amsmath}
\usepackage{multirow}
\biboptions{numbers,sort&compress}
\usepackage{hyperref}

\usepackage{lineno}
\usepackage{float}
\usepackage{bm}
\usepackage{booktabs}
\usepackage{enumitem}

\journal{Nuclear Instruments and Methods A}

\begin{document}

\begin{frontmatter}



\title{A Monte Carlo Method for Rayleigh Scattering in Liquid Detectors}

\author[label1]{Miao Yu}
\author[label2]{Wenjie Wu}
\author[label3]{Yayun Ding}
\author[label4]{Qian Liu}
\author[label1]{Feng Ren\corref{aaa}} 
\ead{fren@whu.edu.cn}
\author[label1]{Zhenyu Zhang}
\author[label1]{Xiang Zhou\corref{bbb}}
\ead{xiangzhou@whu.edu.cn}
\affiliation[label1]{organization={Hubei Nuclear Solid Physics Key Laboratory, School of Physics and Technology},
           addressline={Wuhan University}, 
            city={Wuhan},
            postcode={430072}, 
            state={Hubei},
            country={China}}
\affiliation[label2]{organization={Department of Physics and Astronomy}, 
            addressline={University of California at Irvine},
            cite={Irvine},
            postdoc={92697},
            state={California},
            country={USA}}
\affiliation[label3]{organization={Institute of High Energy Physics},
           addressline={Chinese Academy of Sciences}, 
            city={Beijing},
            postcode={100049}, 
            country={China}}
\affiliation[label4]{organization={School of Physics},
           addressline={University of Chinese Academy of Sciences}, 
            city={Beijing},
            postcode={100049}, 
            country={China}}
\cortext[aaa]{Corresponding Author}
\cortext[bbb]{Corresponding Author}

\begin{abstract}
A new Monte Carlo method has been implemented to describe the angular and polarization distributions of anisotropic liquids, like water and linear alkylbenzene, by considering orientational fluctuations of polarizability tensors. The scattered light of anisotropic liquids is depolarized with an angular distribution of $1+(1-\rho_v)/(1+3\rho_v)\cos^2\theta$, which is modified by the depolarization ratio $\rho_v$. A standalone experiment has validated the simulation results of LAB. The new method can provide more accurate knowledge on light propagation in large liquid detectors, which is beneficial to the developments of reconstruction for detectors.
\end{abstract}


\end{frontmatter}


\section{Introduction} \label{Sec:Intro}
Nowadays, large scale liquid detectors have been widely used in neutrino experiments. One category is water-based Cherenkov detectors such as Super Kamiokande~\cite{Super-Kamiokande:1998kpq} and SNO~\cite{SNO:1999crp}, which consist of large volumes of water (heavy water) surrounded by phototubes that collect Cherenkov radiation by charged particles. The characteristic Cherenkov rings reveal the particle information like direction and energy~\cite{Shiozawa:1999sd}. The other type usually uses organic liquid scintillator (LS) as the target materials. The latter is beneficial to the reactor and solar neutrino exploration like in Daya Bay~\cite{DayaBay:2012fng}, Borexino~\cite{Borexino:2008gab} and the upcoming experiment JUNO~\cite{JUNO:2015zny}.

Rayleigh scattering is an important optical process for light propagation in large liquid detectors~\cite{Rayleigh1881}. Light paths of photons will be modified by scattering which makes both the hit patterns and time distributions different. Depolarization of scattered photons is an important property of anisotropic molecules. The common recipes of liquid detectors mentioned above like water and linear alkylbenzene (LAB) are optically anisotropic fluids, and the depolarization ratios $\rho_v$ have already been measured~\cite{Kratohvil1965, Liu2015, Zhou:2015fwa,Wurm:2010ad}. The magnitude of Rayleigh scattering is supposed to be increased for anisotropic liquids due to the Cabannes correction~\cite{Coumou1964} and related measurements have already been discussed for LAB in Ref.~\cite{Liu2015, Zhou2015}. Besides, the angular distribution has also been modified by the depolarization ratio as $1+(1-\rho_v)/(1+3\rho_v)\cos^2\theta$~\cite{Dawson1941}, which means more photons deviate from the original directions. For large water-based Cherenkov detectors, it is expected that the Cherenkov ring patterns are more smeared which brings concerns to the reconstruction. Therefore, the implementations of Rayleigh scattering of anisotropic liquids in simulation are helpful for the understanding of detector response. However, current common Monte Carlo toolkits like Geant4~\cite{GEANT4:2002zbu}, have not considered Rayleigh scattering of anisotropic liquids yet. In the version Geant4.10.06, both the default implementation \textsl{G4OpRayleigh} class and two Rayleigh scattering classes in the Livermore Model~\cite{Allison:2016lfl} compute the Rayleigh scattering based on the angular distribution of isotropic molecules~\cite{Depaola2003}. In this work, we have implemented a new Monte Carlo method for Rayleigh scattering in liquids which is generally applicable to both isotropic and anisotropic molecules, and has the capacity to reproduce correct angular and polarization distributions.

In this paper, we introduce the theory of Rayleigh scattering of anisotropic molecules in Section~\ref{Sec:RayScat}. Details of Monte Carlo implementations have been presented in Section~\ref{Sec:MCSim}. A brief experiment has been built for validation and the experimental setup has been described in Section~\ref{Sec:Exp}. In Section~\ref{Sec:Results} we present some test results to exhibit the validity of our Monte Carlo simulation. Finally we give a brief summary in Section~\ref{Sec:Summary}.

\section{Rayleigh scattering of anisotropic molecules} \label{Sec:RayScat}
It has been found that depolarization exists in Rayleigh scattering light of many liquids, like water, benzene, LAB and so on\cite{RCC1964,Kratohvil1965, Liu2015, Wurm:2010ad, Cummins1965}. The dominant origin of depolarization for optically anisotropic molecules is orientational fluctuations. 
The probability distribution function of scattered light $f_s$ is approximately proportional to the field correlation functions with only the ``self'' terms~\cite{Cummins1996},
\begin{eqnarray}
    f_s(t) \propto \left< \sum_j \left[\bm{\epsilon}_s\cdot\bm{\alpha}_j(t)\cdot\bm{\epsilon}_i\right]\left[\bm{\epsilon}_s\cdot\bm{\alpha}_j(0)\cdot\bm{\epsilon}_i\right]\times\exp{\left\{i\mathbf{k}\cdot\left[\mathbf{r}_j(t)-\mathbf{r}_j(0)\right]\right\}}\right>,
    \label{Eq:ScatLightIntensity}
\end{eqnarray}
where $\bm{\epsilon}_i$ and $\bm{\epsilon}_s$ are the polarization unit vectors of incident light and scattered light respectively, $\mathbf{k}=\mathbf{k}_s-\mathbf{k}_i$ is the momentum transferred, $\mathbf{r}_j(t)$ is the position of molecule $j$ at time $t$, and $\bm{\alpha}_j(t)$ refers to the polarizability tensor of molecule $j$ at time $t$. For isotropic molecules, the polarizability tensor $\bm{\alpha}$ degenerates to a scalar which induces the light intensity vanishes when $\bm{\epsilon}_i \perp \bm{\epsilon}_s$, thus the scattered light is totally polarized. And the randomly orientated anisotropic molecules lead to the depolarization property of scattered photons. Besides, it is also the major causes of the broaden scattering spectra, so called ``Rayleigh wings"~\cite{BenReuven1969}.

For anisotropic molecules, the equivalent form of axially symmetric polarizability tensor in the principle-axes frame is like Eq.~\eqref{Eq:symAlpha},
\begin{equation}
    \bm{\alpha} = 
    \begin{pmatrix}
    \alpha_\parallel & 0 & 0 \\
    0 & \alpha_\perp & 0 \\
    0 & 0 & \alpha_\perp \\
    \end{pmatrix}.
    \label{Eq:symAlpha}
\end{equation}
And a same relation can be derived as in Ref.~\cite{Cummins1996},
\begin{eqnarray}
    \rho_v \equiv \frac{H_v}{V_v} = \frac{3(\alpha_\parallel-\alpha_\perp)^2}{5(\alpha_\parallel+2\alpha_\perp)^2 + 4(\alpha_\parallel-\alpha_\perp)^2},
    \label{Eq:rhov_polarizability}
\end{eqnarray}
where $H_v$ refers to the light intensity of horizontally polarized scattered light with perpendicularly polarized incident light, and $V_v$ obeys the same syntax. And the perpendicular depolarization ratio $\rho_v$ is a directly measurable physical quantity.

\section{Sampling procedures} \label{Sec:MCSim}
The implementations of Monte Carlo simulation based on discussions in Section~\ref{Sec:RayScat} to sample the angular and polarization distributions of Rayleigh scattered photons are presented in this section. The slight wavelength shifts of the scattered photons are ignored in this work. Compared to the range of molecular correlations, the wavelength of radiation is often much greater so that the exponential part in Eq.~\eqref{Eq:ScatLightIntensity} is approximately equal to one.
Therefore, the simplification form of Eq.~\eqref{Eq:ScatLightIntensity} can be rewritten as 
\begin{eqnarray}
    f_s \propto \left< \left[\bm{\epsilon}_s\cdot\bm{\alpha'}_j\cdot \bm{\epsilon}_i\right]^2\right>,
    \label{Eq:ScatLightIntensity_Simplified}
\end{eqnarray}
where $\bm{\alpha'}$ is an arbitrarily oriented polarizability tensor of molecule $j$ due to fluctuations.
\begin{figure*}
	\centering
	\includegraphics[width=\textwidth]{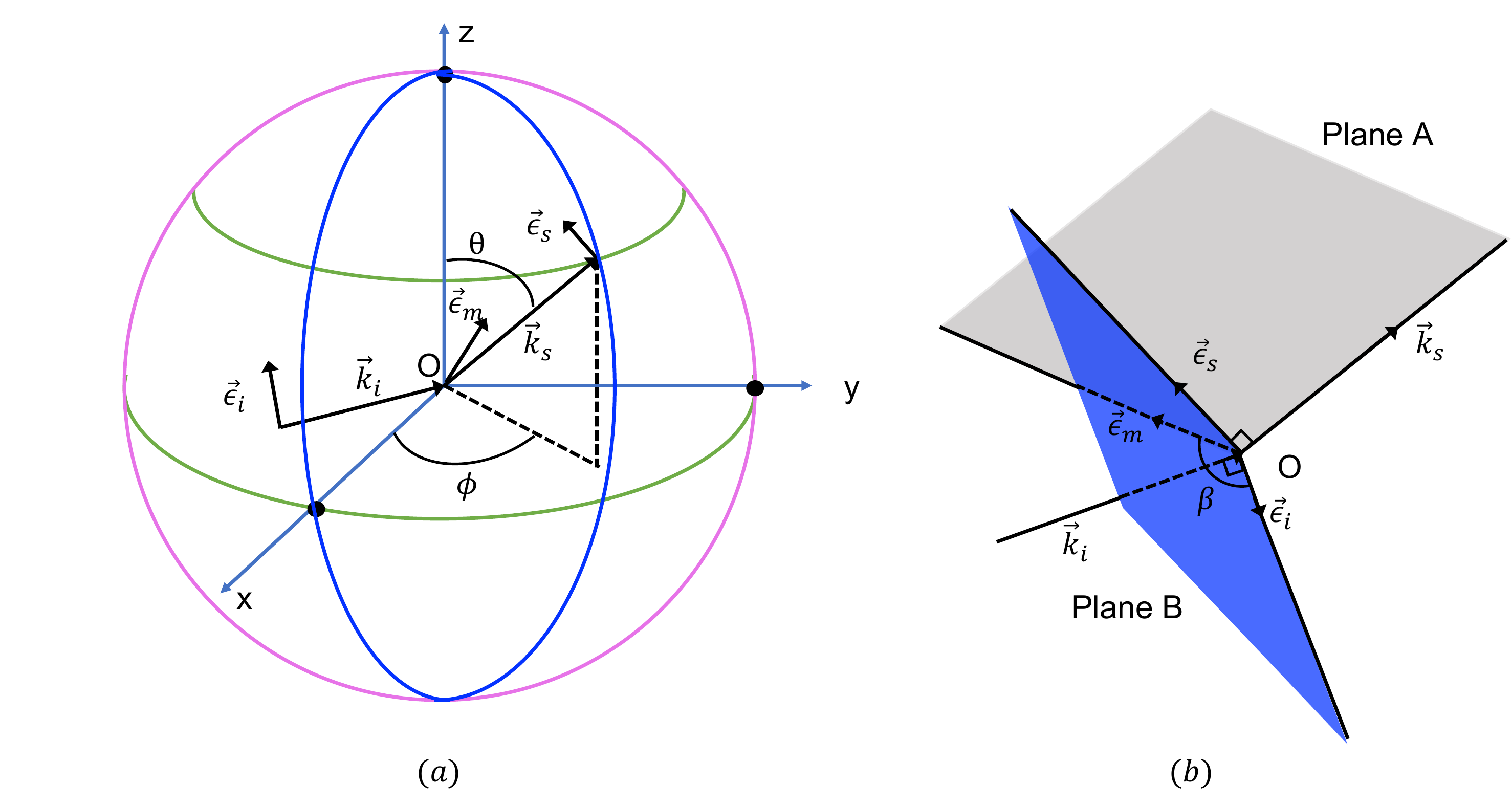}
	\caption{(a) Rayleigh scattering at origin $O$. $\bm\epsilon_i$ and $\mathbf k_i$ are the momentum vector and polarization unit vector of the incident light, while $\bm\epsilon_s$ and $\mathbf k_s$ are the corresponding vectors of the scattered light. $\theta$ and $\phi$ are the polar angle and azimuthal angle of $\mathbf k_s$. $\bm\epsilon_i$ is modified as $\bm\epsilon_m$ by the polarizability tensor. (b) The polarization unit vector $\bm\epsilon_s$ is required at the plane A which is decided by $\bm\epsilon_m$ and $\mathbf k_s$ and perpendicular to $\mathbf k_s$ simultaneously. $\beta$ is the angle between $\bm\epsilon_i$ and $\bm\epsilon_s$ in the plane B.}
	\label{Fig:Scatter}
\end{figure*}

According to Eq.~\eqref{Eq:ScatLightIntensity_Simplified}, the implementations of Monte Carlo simulation are as follows, and a schematic diagram has been displayed in Fig.~\ref{Fig:Scatter},
\begin{enumerate}
    \item The depolarization ratio $\rho_v$ is required as a material property. Then the polarizability tensor in the principle-axes frame can be calculated based on Eq.~\eqref{Eq:rhov_polarizability} where $\alpha_\parallel$ is set as unit for normalization;
    
    \item To consider the orientational fluctuations, rotation matrices are randomly generated and the polarizability tensor is randomly rotated in the space as
    \begin{equation}
        \bm{\alpha'} = \mathbf{M}  \bm{\alpha} \mathbf{M^{-1}},
        \label{Eq:RotatePolMatrix}
    \end{equation}
    where $\mathbf{M}$ is an arbitrary rotation matrix;
    
    \item The scattering momentum ($\mathbf{k}_s$) direction is randomly sampled firstly;
    \begin{enumerate}
        \item Sample a uniform random $\cos{\theta_r}$ within $\left[ -1, 1\right]$;
        \item Sample a uniform random $\phi_r$ within $\left[ 0, 2\pi \right]$;
    \end{enumerate}
    
    \item The polarization unit vector of the incident light ($\bm{\epsilon}_i$) is then modified by the polarizability tensor,
    \begin{equation}
        \bm \epsilon_m = \frac{\bm{\alpha'} \bm{\epsilon}_i}{|\bm{\alpha'} \bm{\epsilon}_i|},
        \label{Eq:IncPolRatation}
    \end{equation}
    where $\bm{\epsilon}_m$ is the modified polarization unit vector. With a given scattering direction ($\theta_r$, $\phi_r$), the polarization unit vector of the scattered photon $\bm{\epsilon}_s$ is required at the plane decided by $\mathbf{k}_s$ and $\bm{\epsilon}_m$ (the plane A in the plot (b) of Fig.~\ref{Fig:Scatter}) while perpendicular to the scattering momentum direction $\mathbf{k}_s$. Therefore, $\bm{\epsilon}_s$ can be determined by the given ($\theta_r$, $\phi_r$);
    
    \item Once the $\bm\epsilon_s$ is determined, the probability distribution function of the scattering direction can be calculated according to Eq.~\eqref{Eq:ScatLightIntensity_Simplified} as,
    \begin{equation}
        P(\theta_r, \phi_r) = \left[\bm{\epsilon}_s \cdot \bm{\alpha'} \cdot \bm{\epsilon}_i \right]^2 ,
        \label{Eq:ScatPDF}
    \end{equation}

    \item Sample a uniform random $P_r$ within $\left[ 0, 1\right]$;
  
  \item Compare $P_r$ with $P(\theta_r, \phi_r)$ (Eq.~\eqref{Eq:ScatPDF}), if $P_r \leq P(\theta_r, \phi_r)$ accept the scattering angle $\theta = \theta_r$ and $\phi = \phi_r$, else restart.
    
\end{enumerate}

The sampling procedures presented above are independent with the coordinate system. All the calculations can be applied in the laboratory reference frame and no reference transformations are required.

\section{Experimental validation} \label{Sec:Exp}
\begin{figure*}
	\centering
	\includegraphics[width=0.7\textwidth]{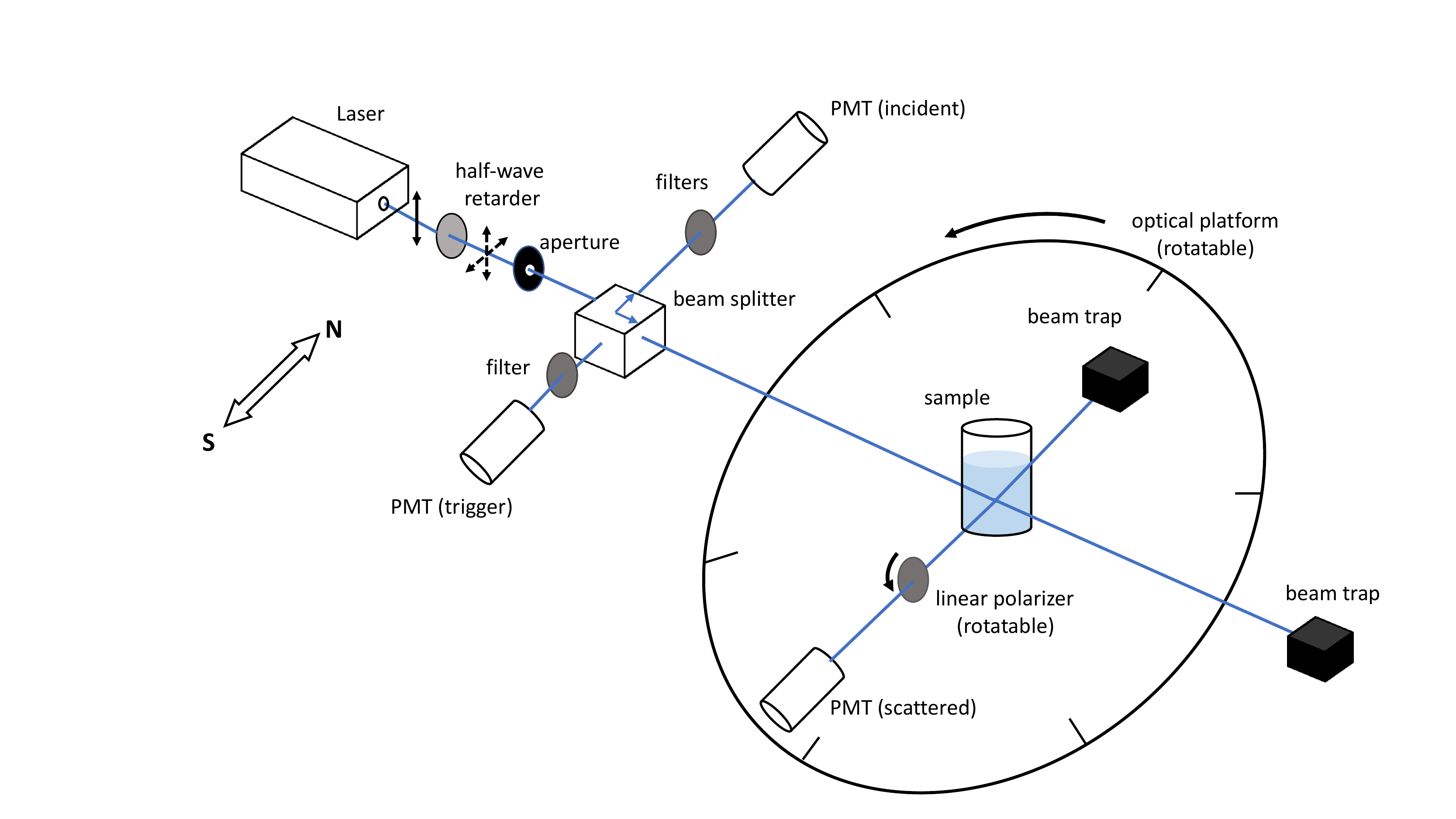}
	\caption{Sketch diagram of the experimental setup.}
	\label{Fig:setup}
\end{figure*}

A bench experiment has been performed to measure the polarization of scattering light of LAB. The experimental setup is schematically shown in Fig.~\ref{Fig:setup}. The whole system was placed in a dark room with stable temperature control at $21^{\circ}$C. The polarization direction of the laser beam was perpendicular to the scattering plane, and horizontally polarized light can be obtained by placing a half-wave retarder and a polarizer after the light source. The wavelength of the polarized light from the laser source (Pico-Quant LDH pulsed diode) was centered at $432\pm0.3$ nm. The incident light was collimated with an aperture and then split into two directions by a beam splitter (Thorlabs CCM1-BS013). One beam was split into the direction perpendicular to the original propagation direction and used as the incident light monitor. The remained beam entered into the LAB samples in a cylindrical quartz cuvette and was scattered with some probability. The collection of a small fraction of reflected light on the beam splitter provided triggers for data-taking. The whole scattering detection part was positioned on an rotatable optical platform for multiple scattering direction measurements. The depolarization components in scattered light were measured by rotating a linear polarizer at the scattering path. Two beam traps were positioned at the end of light path for stray light suppression.

The photon detectors used in the experiment were the photomultiplier tubes (PMTs). Because the Rayleigh scattering length of LAB is relatively long, the scattered light intensity was weak so PMTs worked in the single photon counting mode. The incident moniter beam was relative intense, thus series of filters were used for light attenuation. All PMTs were aligned into south-north orientation to avoid effects from the geomagnetic field. Signals coincident with triggers were recorded by a CAEN Mod. N1145 quad scaler and preset counter. The experiment benefited from the high efficiency data-taking strategy for no waveform sampling.

\section{Simulation results and validation} \label{Sec:Results}
Detailed Monte Carlo simulations for Rayleigh scattering were performed and compared with both the theoretical predictions and the experimental results. For simplification, during the simulation the scattering position was set at the origin and the momentum vector of incident photons was fixed as $\mathbf{k}_i = \left[ 0, 0, E\right]$ where $E$ was the photon energy. Perpendicularly polarized ($\bm\epsilon_i = \left[ 1, 0, 0\right]$), horizontally polarized ($\bm\epsilon_i = \left[ 0, 1, 0\right]$) and unpolarized incident light have been studied respectively. The generation of scattered photons followed the procedures described in Section~\ref{Sec:MCSim}.

\subsection{Angular distributions of scattered light intensity} \label{SubSec:AngDisTest}

\begin{figure*}
	\centering
	\includegraphics[width=\textwidth]{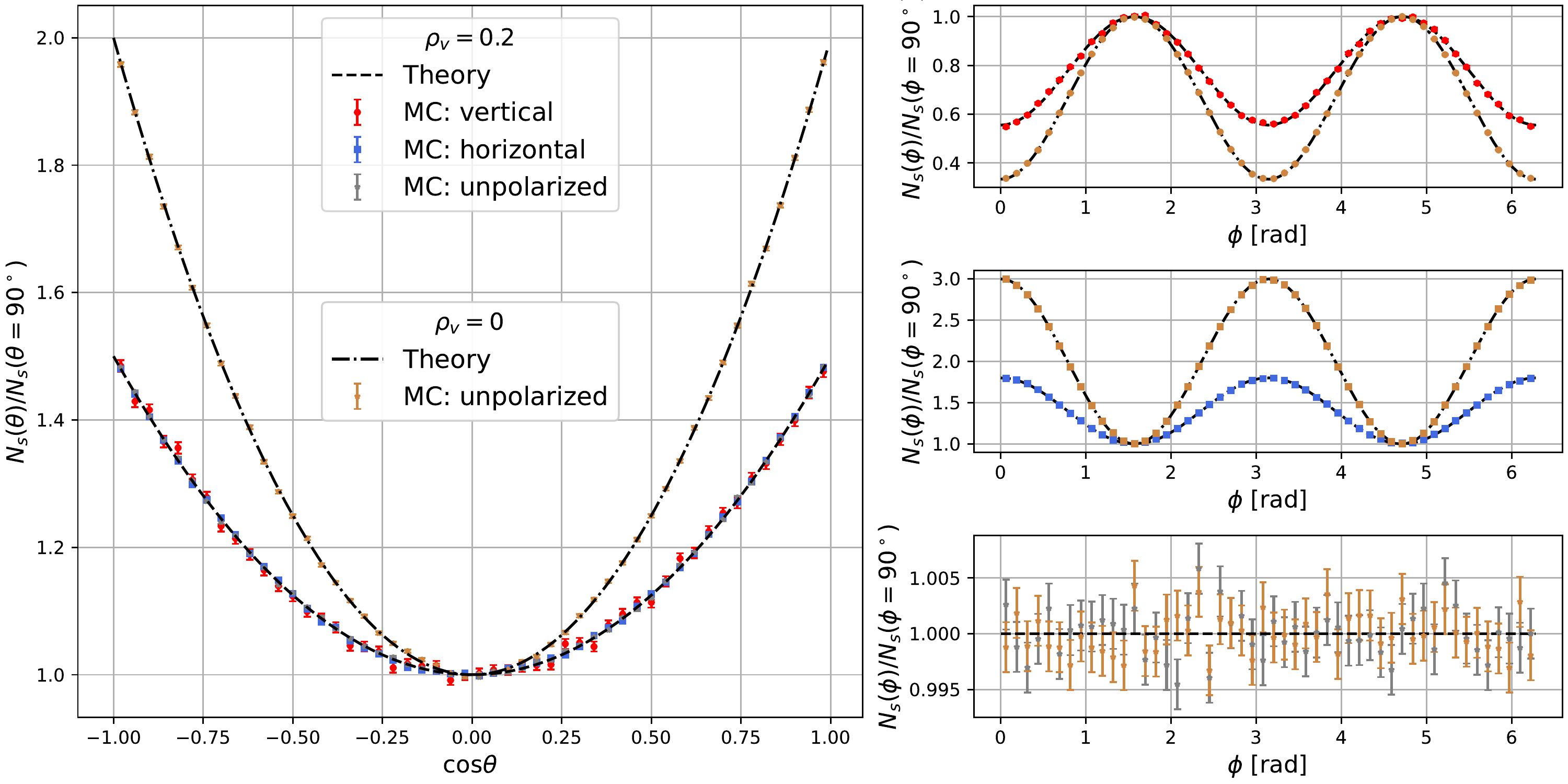}
	\caption{Angular distributions of the scattered light intensity for three different cases. Red circle markers: perpendicularly polarized incident light; blue square markers: horizontally polarized incident light; gray star markers: unpolarized incident light; black dashed lines: theoretical predictions. The y-axis is the relative number of scattered photons compared with $90^\circ$. Left: $\cos\theta$ distributions of scattered photons. Brown markers are simulation results of isotropic molecules which are also consistent with theoretical predictions for comparison. All three cases obey a same distribution and have good agreements with theoretical predictions. Right: azimuthal ($\phi$) distributions of three different polarized incident light cases respectively. }
	\label{Fig:AngDis1D}
\end{figure*}

For anisotropic molecules, the angular distributions of scattered photons can be derived as~\cite{Kerker2013,Zhou2015},
\begin{align}
    I_s(\theta) &= I_s(\theta = 90^\circ)\left( 1 + \frac{1-\rho_v}{1+3\rho_v }\cos^2\theta \right), \label{Eq:theta_dist} \\
    I_s(\phi) &= I_s(\phi = 90^\circ)\left( 1 - \frac{2-2\rho_v}{3+3\rho_v }\cos^2\phi \right). \label{Eq:phi_dist} \\
\end{align}
The one dimensional angular distributions ($\cos\theta$, $\phi$ respectively) from Monte Carlo simulation are shown in Fig.~\ref{Fig:AngDis1D} as markers, which are consistent with the theoretical predictions. The perpendicular depolarization ratio $\rho_v$ was set as $0.2$ for the anisotropic cases in simulation. The three polarization cases have the same $\cos\theta$ distribution because the $\phi$ integration averages the polarization of incident light out. The $\cos\theta$ distribution of isotropic molecules is also displayed, and it's verified that by setting $\rho_v=0$, the Monte Carlo implementations are also compatible with isotropic liquids. And it is noticed that for anisotropic liquids, the scattering directions deviate from the original directions to a larger extent. Complementary azimuthal modulation curves have been obtained for the two polarized incident light cases while the unpolarized incident light case has a flat $\phi$ distribution as expected. 

\begin{figure*}
	\centering
	\includegraphics[width=\textwidth]{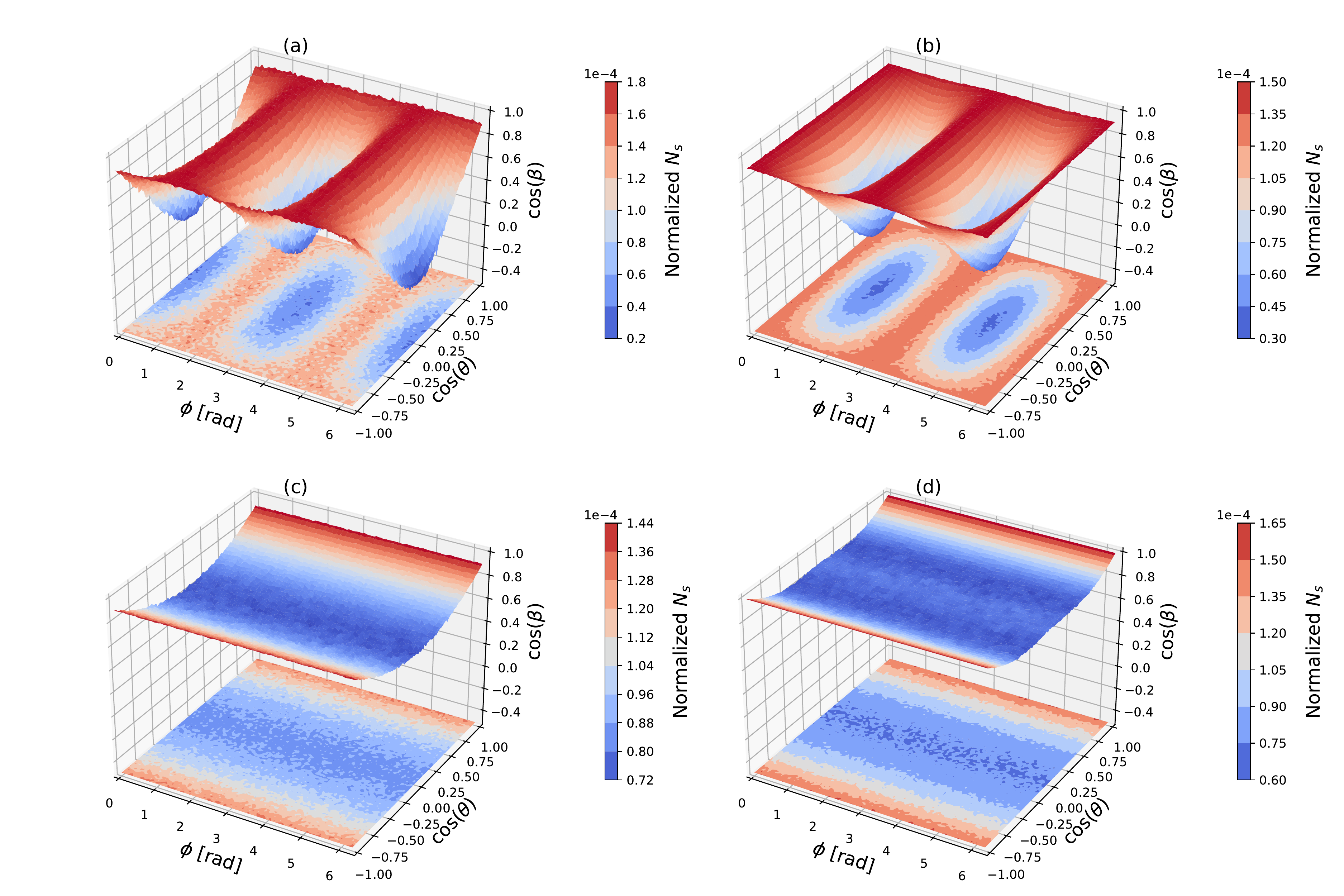}
	\caption{$\cos\beta$ with respect to $\cos\theta$ and $\phi$, and the two dimensional $\cos\theta$-$\phi$ distributions are projected below, where the detected scattered photon numbers $N_s$ are normalized. (a) perpendicularly polarized incident light ($\rho_v=0.2$); (b) horizontally polarized incident light ($\rho_v=0.2$); (c) unpolarized incident light ($\rho_v=0.2$); (d) unpolarized incident light ($\rho_v=0$).}
	\label{Fig:AngDis3D}
\end{figure*}

Furthermore in Fig.~\ref{Fig:AngDis3D} we show the angle between $\bm{\epsilon}_i$ and $\bm{\epsilon}_s$, which is defined as 
\begin{equation}
    \cos\beta = \bm{\epsilon}_i \cdot \bm{\epsilon}_s,
    \label{Eq:beta}
\end{equation}
and shown in the plane B in the plot (b) of Fig.~\ref{Fig:Scatter}, with respect to $\cos\theta$ and $\phi$ from simulation. The two dimensional $\cos\theta-\phi$ distributions have been projected below respectively. A relatively larger density of scattered photons are yielded where the polarization vector of the scattered light is more parallel to that of the incident light.

\subsection{Depolarization in different scattering directions} \label{SubSec:DepolarTest}
\begin{figure*}
	\centering
	\includegraphics[width=\textwidth]{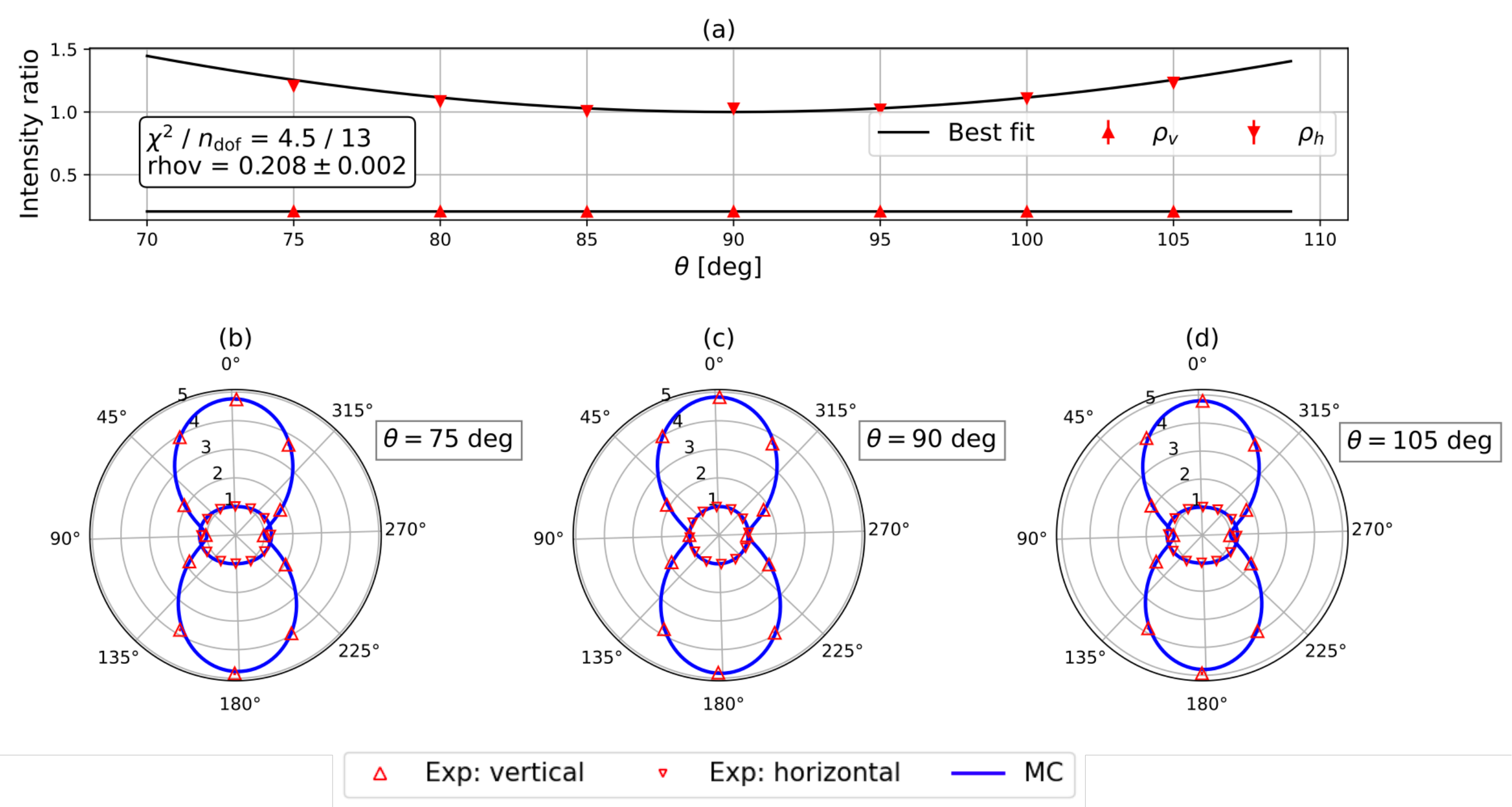}
	\caption{(a) Combined fitting of $\rho_h$ and $\rho_v$ of LAB measurements. Red markers are measured data of LAB, black lines are the best fits. Polar plots (b)-(d) of scattered light intensity in different polarization directions with both perpendicularly and horizontally polarized incident light at $75^\circ$, $90^\circ$ and $105^\circ$ respectively, where $H_v$ and $V_h$ are scaled as unit for visualization. Red markers are experimental data and blue lines are Monte Carlo simulation, and good agreements have been obtained. Results at $75^\circ$ and $105^\circ$ are symmetry as expected. }
	\label{Fig:rho+polar}
\end{figure*}

During the experiment, the depolarization components of scattered light of LAB at different $\theta$ angles were measured by rotating the polarizer and detecting the intensity of scattered light. Total $12$ polarization angles have been measured at each scattering $\theta$ angle, where $\theta$ spans from $75^\circ$ to $105^\circ$ per $5^\circ$. And the measurement results of LAB for both perpendicularly and horizontally polarized incident light have been shown as red markers in plots (b)-(d) of Fig.~\ref{Fig:rho+polar}, where $H_v$ and $V_h$ are scaled as one for visualization.

According to the relationships in Eq.~\eqref{Eq:theta_dist}-~\eqref{Eq:phi_dist}, the angular distributions of perpendicular and horizontal depolarization ratios ($\rho_v$, $\rho_h$) can be deduced as,
\begin{align}
    \rho_v(\theta) &= \mathrm{const}, \label{Eq:rhov_theta} \\
    \rho_h(\theta) &\equiv \frac{H_h}{V_h} = \cos^2\left(\theta\right)\frac{1}{\rho_v} + \sin^2\left(\theta\right). \label{Eq:rhoh_theta}
\end{align}
Thus, a combined fitting of the experiment results gives a measurement value as $\rho_v = 0.208 \pm 0.002$ in the plot (a) of Fig.~\ref{Fig:rho+polar} for the LAB samples.

By setting the measured $\rho_v$ as the material property, simulations of Rayleigh scattering of LAB have been performed. The configurations of the simulation kept the same with the experimental layouts, where the scattered radiation was detected by an ideal polarimetry at the yOz plane. The depolarization components from the Monte Carlo have been displayed as blue lines in plots (b)-(d) of Fig.~\ref{Fig:rho+polar}. Good agreements between experimental data and Monte Carlo simulations validate the reliability of our codes. Also the $\theta$-symmetry has been revealed in results at $75^\circ$ and $105^\circ$ as our predictions.

\section{Summary and Discussions} \label{Sec:Summary}
In this work, we present a comprehensive Monte Carlo method to sample Rayleigh scattering light. By considering orientational fluctuations of polarizability tensors, the method is capable to reproduce the angular and polarization status of the scattered radiation of anisotropic liquids, which are consistent with theoretical predictions. Meanwhile, by setting the depolarization ratio as zero, it is naturally compatible with the results of isotropic molecules. Besides, a bench experiment has been built to test Rayleigh scattering of LAB. The Monte Carlo method can also yield consistent results with the experimental data so that the validity of the implementations was confirmed furthermore.

The new Monte Carlo method provides a more proper consideration of light propagation in large scale liquid detectors. For isotropic molecules, the scattered light obeys the angular distribution $1+\cos^2\theta$ where forward and backward scattering are preferred. While for anisotropic liquids like water, the Eq.~\eqref{Eq:theta_dist} indicates a flatter distribution in all directions. Potential concerns could be brought especially to the large scale water-based Cherenkov detectors, where the smeared Cherenkov ring patterns might challenge the previous reconstruction algorithms. For the next generation of large scale water-based detector Hyper-Kamiokande, which has a cylindrical water tank of $71$ m depth and $68$ m diameter~\cite{Hyper-Kamiokande:2016dsw}, such effects might be remarkable due to the long optical paths. The renewed light propagation in Monte Carlo may help for the understanding of detector response and motivate the developments of more accurate reconstruction algorithms.

\section*{Acknowledgments}
This work has been supported by the Major Program of the National Natural Science Foundation of China (Grant No.11390381).


\begin{thebibliography}{00}

\bibitem{Super-Kamiokande:1998kpq}
Y.~Fukuda \textit{et al.} [Super-Kamiokande],
Phys. Rev. Lett. \textbf{81}, 1562-1567 (1998)
doi:10.1103/PhysRevLett.81.1562
[arXiv:hep-ex/9807003 [hep-ex]].

\bibitem{SNO:1999crp}
J.~Boger \textit{et al.} [SNO],
Nucl. Instrum. Meth. A \textbf{449}, 172-207 (2000)
doi:10.1016/S0168-9002(99)01469-2
[arXiv:nucl-ex/9910016 [nucl-ex]].

\bibitem{Shiozawa:1999sd}
M.~Shiozawa [Super-Kamiokande],
Nucl. Instrum. Meth. A \textbf{433}, 240-246 (1999)
doi:10.1016/S0168-9002(99)00359-9

\bibitem{DayaBay:2012fng}
F.~P.~An \textit{et al.} [Daya Bay],
Phys. Rev. Lett. \textbf{108}, 171803 (2012)
doi:10.1103/PhysRevLett.108.171803
[arXiv:1203.1669 [hep-ex]].

\bibitem{Borexino:2008gab}
G.~Alimonti \textit{et al.} [Borexino],
Nucl. Instrum. Meth. A \textbf{600}, 568-593 (2009)
doi:10.1016/j.nima.2008.11.076
[arXiv:0806.2400 [physics.ins-det]].

\bibitem{JUNO:2015zny}
F.~An \textit{et al.} [JUNO],
J. Phys. G \textbf{43}, no.3, 030401 (2016)
doi:10.1088/0954-3899/43/3/030401
[arXiv:1507.05613 [physics.ins-det]].

\bibitem{Rayleigh1881}
Rayleigh, Lord. 
The London, Edinburgh, and Dublin Philosophical Magazine and Journal of Science 12.73 (1881): 81-101.

\bibitem{Liu2015}
Q.~Liu, X.~Zhou, W.~Huang, Y.~Zhang, W.~Wu, W.~Luo, M.~Yu, Y.~Zheng, L.~Zhou and J.~Cao, \textit{et al.}
Nucl. Instrum. Meth. A \textbf{795}, 284-287 (2015)
doi:10.1016/j.nima.2015.05.032
[arXiv:1504.01001 [physics.ins-det]].

\bibitem{Zhou:2015fwa}
X.~Zhou, Q.~Liu, J.~Han, Z.~Zhang, X.~Zhang, Y.~Ding, Y.~Zheng, L.~Zhou, J.~Cao and Y.~Wang,
Eur. Phys. J. C \textbf{75}, no.11, 545 (2015)
doi:10.1140/epjc/s10052-015-3784-z
[arXiv:1504.00986 [physics.chem-ph]].


\bibitem{Wurm:2010ad}
M.~Wurm, F.~von Feilitzsch, M.~Goeger-Neff, M.~Hofmann, T.~Lachenmaier, T.~Lewke, T.~Marrod\'an Undagoitia, Q.~Meindl, R.~M\"ollenberg and L.~Oberauer, \textit{et al.}
Rev. Sci. Instrum. \textbf{81}, 053301 (2010)
doi:10.1063/1.3397322
[arXiv:1004.0811 [physics.ins-det]].


\bibitem{Kratohvil1965}
Kratohvil, J. P., M. Kerker, and L. E. Oppenheimer. 
The Journal of Chemical Physics 43.3 (1965): 914-921.


\bibitem{Coumou1964}
Coumou, D. J., E. L. Mackor, and J. Hijmans. 
Transactions of the Faraday Society 60 (1964): 1539-1547.


\bibitem{Zhou2015}
X.~Zhou, Q.~Liu, M.~Wurm, Q.~Zhang, Y.~Ding, Z.~Zhang, Y.~Zheng, L.~Zhou, J.~Cao and Y.~Wang,
Rev. Sci. Instrum. \textbf{86}, no.7, 073310 (2015)
doi:10.1063/1.4927458
[arXiv:1504.00987 [physics.ins-det]].

\bibitem{Dawson1941}
Dawson, L. H., and E. O. Hulburt. 
JOSA 31.8 (1941): 554-558.


\bibitem{GEANT4:2002zbu}
S.~Agostinelli \textit{et al.} [GEANT4],
Nucl. Instrum. Meth. A \textbf{506}, 250-303 (2003)
doi:10.1016/S0168-9002(03)01368-8

\bibitem{Allison:2016lfl}
J.~Allison, J.~Apostolakis, S.~B.~Lee, K.~Amako, S.~Chauvie, A.~Mantero, J.~I.~Shin, T.~Toshito, P.~R.~Truscott and T.~Yamashita, \textit{et al.}
Nucl. Instrum. Meth. A \textbf{835}, 186-225 (2016)
doi:10.1016/j.nima.2016.06.125

\bibitem{Depaola2003}
Depaola, G. O. 
Nuclear Instruments and Methods in Physics Research Section A: Accelerators, Spectrometers, Detectors and Associated Equipment 512.3 (2003): 619-630.

\bibitem{RCC1964}
Leite, Rogerio CC, Robert S. Moore, and Sergio PS Porto. 
The Journal of Chemical Physics 40.12 (1964): 3741-3742.

\bibitem{Cummins1965}
Cummins, Herman Z., and Robert W. Gammon. 
Applied Physics Letters 6.8 (1965): 171-173.

\bibitem{Cummins1996}
Cummins, H. Z., et al. 
Physical Review E 53.1 (1996): 896.

\bibitem{BenReuven1969}
Ben‐Reuven, A., and N. D. Gershon. 
The Journal of Chemical Physics 51.3 (1969): 893-902.


\bibitem{Kerker2013}
Kerker, Milton. The scattering of light and other electromagnetic radiation: physical chemistry: a series of monographs. Vol. 16. Academic press, 2013.

\bibitem{Hyper-Kamiokande:2016dsw}
 [Hyper-Kamiokande],
``Hyper-Kamiokande Design Report,''
KEK-PREPRINT-2016-21.

\end{thebibliography}
\end{document}